 \documentstyle[preprint,aps]{revtex}

\draft
\newcommand{\be}{\begin{equation}}
\newcommand{\ee}{\end{equation}}
\newcommand{\ben}{\begin{eqnarray}}
\newcommand{\een}{\end{eqnarray}}

\newcommand{\iii}{\'{\i}}

\begin{document}

\draft
\title{On the Entanglement Properties of Two-Rebits Systems.}
\author{J. Batle$^1$, A. R. Plastino$^{1,\,2,\,3}$, M. Casas$^1$\footnote{E-mail:
dfsmca0@uib.es, corresponding author.}, and A. Plastino$^{2,\,3}$}

\address {$^1$Departament de F\iii sica, Universitat de les Illes Balears,
07071 Palma de Mallorca, Spain \\
$^2$National University La Plata, C.C. 727, 1900 La Plata, Argentina \\
  $^3$Argentina's National Research Council (CONICET) }

\date{\today}

\maketitle
 \begin{abstract}

 Following the recent work of Caves, Fuchs, and Rungta
 [Found. of Phys. Lett. {\bf 14} (2001) 199], we
 discuss some entanglement properties of two-rebits
 systems. We pay particular attention to the relationship
 between entanglement and purity. In particular, we determine
 (i) the probability densities for finding pure and mixed
 states with a given amount of entanglement, and (ii)
 the mean entanglement of two-rebits states as a function
 of the participation ratio.

\vskip 5mm
 Pacs: 03.67.-a; 89.70.+c; 03.65.-w

\vskip 5mm


\end{abstract}
\vspace{.5cm}

\maketitle

\newpage

\section{Introduction}

It has been recently pointed out by  Caves, Fuchs,  and Rungta
\cite{CFR01} that real quantum mechanics (that is, quantum
mechanics defined over real vector spaces
\cite{S60,GPRS61,E86,W02}) provides an interesting foil theory
whose study may shed some light on which aspects of quantum
entanglement are unique to standard quantum theory, and which ones
are more generic over other physical theories endowed with the
phenomenon of entanglement.

 Nowadays there is
general consensus on the fact that the phenomenon of entanglement
is one of the most fundamental and non-classical features
exhibited by quantum systems \cite{LPS98}. Quantum entanglement is
the basic resource required to implement several of the most
important processes studied by quantum information theory
\cite{LPS98,WC97,W98,BEZ00,AB01,GD02}, such as  quantum
teleportation \cite{BBCJPW93}, and superdense coding \cite{BW93}.
A state of a composite quantum system constituted by subsystems
$A$ and $B$ is called ``entangled" if it can not be represented as
a convex linear combination of product states. In other words, the
density matrix $\rho^{AB}$ represents an entangled state if it can
not be expressed as

\be
\label{sepa} \rho^{AB} \, = \, \sum_k \, p_k \, \rho^{A}_k
\otimes \rho^{B}_k,
\ee

\noindent with $0\le p_k \le 1$ and $\sum_k p_k =1$. On the
contrary, states of the form (\ref{sepa}) are called separable.
The above definition is physically meaningful because entangled
states (unlike separable states) cannot be prepared locally by
acting on each subsystem individually \cite{P93}. The entanglement
of formation provides a natural quantitative  measure of
entanglement with a clear physical motivation \cite{BDSW96,WO98}.

   In standard quantum mechanics the simplest systems exhibiting
  the phenomenon of entanglement are two-qubits systems.
  They play a fundamental role in Quantum Information Theory.
  It should be stressed that the concomitant space of (mixed)
  two-qubits states is $15$-dimensional and its properties are
  not trivial. An explicit expression for the entanglement of
  formation of a two-qubits state $\rho$ has been found by Wootters
  \cite{WO98}. Wootters' celebrated formula has allowed for a
  systematic survey of the entanglement properties of the space
  of two-qubits states \cite{ZHS98,Z99,MJWK01,IH00}.

  Within quantum mechanics defined over real vector spaces, the
  most basic kind of composite systems are two-rebits systems.
  Rebits are systems whose (pure) states are described by
  normalized vectors in a two dimensional real vector space.
  A rebit may be regarded as the simplest possible quantum
  object \cite{W02}. An explicit expression for the
  entanglement of formation of arbitrary states of
  two-rebits has been obtained by Caves, Fuchs and
  Rungta \cite{CFR01}.

  The aim of the present work is to explore numerically
 the entanglement properties of two-rebits systems. We
 pay particular attention to the relationship between
 entanglement and purity.

  The paper is organized as follows. In section II we review the
  CFR expression for the entanglement of formation of arbitrary
  two-rebits states, and discuss some of its immediate
  consequences. The relationships, for two-rebits systems,
  between the amount of entanglement and the degree of mixture
  are investigated in sections III. Finally, some conclusions
  are drawn in section IV.

\section{The CFR Formula and Some of its Consequences}

Caves, Fuchs,  and Rungta formula  for the entanglement of
formation of a two-rebits state $\rho$ reads \cite{CFR01}

\be \label{renta1} E[\rho] \, = \, h\left(
\frac{1+\sqrt{1-C^2}}{2}\right), \ee

\noindent where

\be \label{renta2} h(x) \, = \, -x \log_2 x \, - \,
(1-x)\log_2(1-x), \ee

\noindent and the concurrence $C$ is given by

\be \label{renta3} C[\rho] \, = \, \mid \! {\rm tr} (\tau) \! \mid
\, = \,
 \mid \! {\rm tr} (\rho \, \sigma_y \otimes \sigma_y) \! \mid.
\ee

\noindent The above expression has to be evaluated by recourse to
the matrix elements of $\rho$ computed  with respect to the
product basis, $\mid \! i,j \rangle = \mid i \rangle \! \mid j
\rangle, \,\, i,j=0,1$.

We are also going to need a quantitative measure of mixedness. There are
several measures of the degree of mixture that can be useful within the present
context. The von Neumann measure

  \be \label{slog}
  S \, = \,- \, Tr \left( \rho \ln \rho \right),
  \ee

  \noindent
  is important because of its relationship with the thermodynamic
  entropy. On the other hand, the so called participation
  ratio,

  \be \label{partrad}
  R(\rho) \, = \, \frac{1}{Tr(\rho^2)},
  \ee

  \noindent
  is particularly convenient for calculations \cite{ZHS98,MJWK01}.

  A remarkable property of two-rebits states, which transpires
  immediately
  from the CFR expressions (\ref{renta1}-\ref{renta3}), is that
  their square concurrence (and, consequently, their entanglement
  of formation) are completely determined by the expectation value
  of one single observable, namely, $\sigma_y \otimes \sigma_y$.
  On the contrary, it has been recently proved that there
  is no observable (not even for pure states) whose sole
  expectation value constitutes enough information to determine
  the entanglement of a two-qubits state \cite{SH00}.
  The operator $\sigma_y \otimes \sigma_y$ has eigenvalues $1$ and
  $-1$, both two-fold degenerated. Let us denote by
  $\mid \phi_{1,2}\rangle  $ the pair of eigenvectors with
  eigenvalue $1$, and $\mid \phi_{3,4}\rangle $ the eigenvectors
  with eigenvalue $-1$, so that

  \be \label{sigsig}
  \sigma_y \otimes \sigma_y \, = \, \sum_{i=1}^2
  \mid \phi_{i}\rangle  \langle \phi_{i}\mid
   \, - \, \sum_{i=3}^4
   \mid \phi_{i}\rangle  \langle \phi_{i}\mid.
  \ee

  \noindent
  A notable consequence of the CFR expressions
  (\ref{renta1}-\ref{renta3}) is that there are
  mixed states of two rebits with maximum entanglement
  (that is, with $C^2=1$). For instance all states of
  the form

  \be \label{perro}
  \rho \, = \, p \! \mid \phi_{1}\rangle  \langle \phi_{1}\mid
  \, + \,
   (1-p) \! \mid \phi_{2}\rangle  \langle \phi_{2}\mid,
  \ee

  \noindent
 with $0\le p \le 1$, are maximally entangled. Hence, for
 any participation rate within the range $1\le R \le 2$
 there exist two-rebits states with maximum entanglement.
 We shall return to this point later, when we discuss the
 distribution of general two-rebits states in
 the $(R,C^2)$-plane.

\section{Entanglement vs. Purity for Arbitrary Two-Rebits States.}

\subsection{Measure on the Two-Rebits State Space}

In order to explore numerically the properties of arbitrary two-rebits states,
it is necessary to introduce an appropriate measure $\mu $ on the space ${\cal
S}_R$ of general two-rebits states. Such a measure is needed to compute volumes
within the space ${\cal S}_R$, as well as to determine what is to be understood
by a uniform distribution of states on ${\cal S}_R$. In order to find a natural
measure on ${\cal S}$ we are going to follow a line of reasoning akin to the
one pursued by Zyczkowski {\it et al.} \cite{ZHS98,Z99} in the case of
two-qubits states.

 An arbitrary (pure and mixed) state $\rho$ of
a (real) quantum system described by an $N$-dimensional real
Hilbert space can always be expressed as the product of three
matrices,

\be \label{odot} \rho \, = \, R D[\{\lambda_i\}] R^{T}. \ee

\noindent Here $R$ is an $N\times N$ orthogonal matrix and
$D[\{\lambda_i\}]$ is an $N\times N$ diagonal  matrix whose
diagonal elements are $\{\lambda_1, \ldots, \lambda_N \}$, with $1
\ge \lambda_i \ge 0$, and $\sum_i \lambda_i = 1$.
   The group of orthogonal matrices $O(N)$ is
endowed with a unique, uniform measure $\nu$ \cite{PZK98}. On the
other hand, the simplex $\Delta$, consisting of all the real
$N$-uples $\{\lambda_1, \ldots, \lambda_N \}$ appearing in
(\ref{odot}), is a subset of a $(N-1)$-dimensional hyperplane of
${\cal R}^N$. Consequently, the standard normalized Lebesgue
measure ${\cal L}_{N-1}$ on ${\cal R}^{N-1}$ provides a natural
measure for $\Delta$. The aforementioned measures on $O(N)$ and
$\Delta$ lead then to a natural measure $\mu $ on the set ${\cal
S}_R$ of all the states of our (real) quantum system, namely,

\be \label{memu}
 \mu = \nu \times {\cal L}_{N-1}.
 \ee

 We are going to consider the set of states of a two-rebits
 system. Consequently, our system will have $N=4$.
 All our present considerations are based on the assumption
 that the uniform distribution of states of a two-rebit system
 is the one determined by the measure (\ref{memu}). Thus, in our
 numerical computations we are going to randomly generate
 states of a two-qubits system according to the measure
 (\ref{memu}).

\subsection{Entanglement and Degree of Mixture.}

The relationship between the amount of entanglement and the purity
of quantum states of composite systems has been recently discussed
in the literature \cite{ZHS98,Z99,MJWK01,IH00}.
  The amount of entanglement and the purity of quantum states of composite
  systems exhibit a dualistic relationship. As the degree of
  mixture increases, quantum states tend to have a smaller
  amount of entanglement. In the case of two-qubits systems,
  states with a large enough degree of mixture are always
  separable \cite{ZHS98}.
  To study the relationship between entanglement and mixture
  we need quantitative measures for these two quantities.
  As already mentioned, the entanglement of formation provides a natural
  quantitative  measure of entanglement with a clear physical motivation
   \cite{BDSW96,WO98}.


The continuous line in Fig. 1 depicts the behavior of the mean
entanglement of formation $\langle E \rangle $ of real density
matrices, as given by the CFR formula,  as a function of the
participation ratio $R$. The dashed line in Fig. 1 corresponds to
the mean entanglement of the same matrices, as given by Wootters'
formula. In other words, in Fig. 1 the continuous line describes
the mean entanglement of formation of the real density matrices
when regarded as defined on a real vector space, while the dashed
line describes the entanglement of formation of these same
matrices when they are considered in the context of the standard
complex vector space. We see that the CFR formula always gives,
for the mean entanglement of formation, a value larger than the
one obtained by recourse of the Wootters' expression. In this
respect, our numerical results are fully consistent with the
general arguments provided  in \cite{CFR01}.

   The continuous line in Fig. 2 illustrates the behavior of the mean
entanglement of formation $\langle E \rangle $  of real density matrices (given
by the CFR expression) as a function of the participation ratio $R$. The dashed
line in Fig. 2 shows the behavior of the mean entanglement of formation
$\langle E \rangle $  of complex density matrices (given by Wootters' formula)
as a function of the participation ratio $R$.

The largest eigenvalue  $\lambda_m$ of the density matrix
constitutes a legitimate measure of mixture, in the sense that
states with larger values of $\lambda_m$ can be regarded as
 less mixed. Its extreme values
correspond to (i) pure states (with $\lambda_m =1$) and (ii) the
density matrix $\frac{1}{4}\hat I$ (with $\lambda_m = 1/4$). In
Fig. 3 we depict the mean entanglement $\langle E\rangle$ of all
the two-rebits states with a given value of their maximum
eigenvalue $\lambda_m$, as a function of this last quantity. The
upper line corresponds to the CFR expression and the lower line to
Wootters formula. Notice that, in the case of Wootters formula,
the mean entanglement vanishes for $\lambda_m \le 1/3 $.

 We have also computed numerically the probability $P(E)$ of finding
 a two-rebits state endowed with an amount of entanglement $E$.
In Fig. 4 we compare (i) the distributions associated with two-rebits states
with (ii) the distributions associated with two-qubits states which were
recently obtained by Zyczkowski {\it et al.} \cite{ZHS98,ZS01}. Fig. 4a depicts
the probability $P(E)$ of finding two-qubits states endowed with a given
entanglement $E$ (as computed with Wootters' expression). The solid line
correspond to arbitrary states and the dashed line to pure states. In a similar
way, Fig. 4b exhibits a plot of the probability $P(E)$ of finding two-rebits
states endowed with a given entanglement $E$ (as computed with the CFR
formula). The solid line correspond to arbitrary states and the dashed line to
pure states. Comparing Figs 4a and 4b we find that the distributions $P(E)$
describing arbitrary states (that is, both pure and mixed states) exhibit the
same qualitative shape for both two-qubits and two-rebits states: in the two
cases the distribution $P(E)$ is a decreasing function of $E$. On the contrary,
the distribution $P(E)$ corresponding to pure two-rebits states differs
considerably from the one associated with pure two-qubits states. The
probability distribution $P(E)$ for pure states of two-rebits reaches its
maximum value for separable states ($E=0$), and it is a monotonous decreasing
function of the entanglement of formation $E$. On the contrary, the
distribution corresponding to pure states of two-qubits is an increasing
function of $E$ for low values of the entanglement, and decreases with $E$ for
large enough values of this variable. It adopts its maximum value for an
intermediate value of $E$. The general conclusion that we may draw from Fig. 4
is that the two curves representing the distributions $P(E)$ associated with
(i) pure states and (ii) arbitrary states do not differ, in the case of
two-rebits states,  as much as they do in the case of two-qubits states.

  The distribution $P(E)$ for pure two-rebits states can be
  obtained analytically. Let us write a pure two-rebits state
  in the form

 \be \label{repuro}
\mid \Psi \rangle \, = \,  \sum_{i=1}^4 \, c_i
  \mid \phi_{i}\rangle,
 \ee

 \noindent
 where

 \be \label{esfera4}
 \sum_{i=1}^4 \, c_i^2 \, = \, 1, \,\,\,\,\,\, c_i\in {\cal R}.
 \ee

 \noindent
 The states $( \mid \phi_{i}\rangle, \,\,\, i=1,\ldots, 4$)
 are the eigenstates of the operator $\sigma_y \otimes \sigma_y$,
 in the same order as in equation (\ref{sigsig}). The four real
 numbers $c_i$ constitute the coordinates of a point lying in
 the three dimensional unitary hyper-sphere $S_3$ (which is
 embedded in ${\cal R}^4$). We now introduce on $S_3$ three
 angular coordinates, $\phi_{1}$, $\phi_{2}$, and $\theta $,
 defined by

 \ben \label{triangle}
 c_1 \, &=& \, \cos \theta \cos \phi_1, \cr
 c_2 \, &=& \, \cos \theta \sin \phi_1, \cr
 c_3 \, &=& \, \sin \theta \cos \phi_2, \cr
 c_4 \, &=& \, \sin \theta \sin \phi_2, \,\,\,\,\,0\le \theta < \frac{\pi}{2},
 \,\, 0\le \phi_1,\phi_2 <2\pi.
 \een

\noindent In terms of the above angular coordinates, the
concurrence of the pure state $\mid \!\Psi \rangle $ is given by

\be \label{cetita} C \, = \, \mid \! \langle \sigma_y \otimes
\sigma_y \rangle \! \mid \, = \, \mid \!\cos 2\theta \!\mid. \ee

\noindent The element of volume on the three dimensional
hyper-sphere is $\sin \theta \cos \theta d\theta d\phi_1 d\phi_2
$. Thus, the total volume associated with a small interval
$d\theta $ is

\be \label{tirita} dv=4\pi^2 \sin \theta \cos \theta \, d\theta.
\ee

\noindent Inspection of equations (\ref{cetita}) and
(\ref{tirita}) allows one to deduce that the probability density
$P(C)$ of finding a pure two-rebits state with concurrence $C$ is

\be \label{pcpc}
 P(C) \, = \, \frac{1}{\pi^2} \, \left| \frac{dv}{d\theta} \right|
 \left| \frac{d\theta}{dC} \right| \, = \, 1.
\ee

\noindent The concomitant probability density $P(E)$ of finding a
pure state with entanglement of formation $E$ is then equal to

\be \label{pepe}
 P(E) \, = \, \left| \frac{dE}{dC} \right|^{-1},
\ee

\noindent where $dE/dC$ is to be computed from expressions (\ref{renta1}) and
(\ref{renta2}). It can be verified from equation (\ref{pepe}) that the limit
value of $P(E)$ associated with states of maximum entanglement is $P(E=1)=\ln
2$. This value corresponds to the horizontal line in Figure 4b.

\subsection{Maximum Entanglement Compatible with a Given Degree of Mixture.}

 We are now going to determine which is the maximum entanglement
 $E_m$ that a two-rebits state with a given participation radio $R$
 may have. Since $E$ is a monotonic increasing function of the
 concurrence $C$, we shall find the maximum value of $C$
 compatible with a given value of $R$. In order to solve the
 concomitant variational problem (and bearing in mind that
 $C = \mid \!\! \langle \sigma_y \otimes \sigma_y \rangle \!\! \mid$ ),
 let us {\it first} find the state that extremizes
  ${\rm Tr} (\rho^2)$ under the constraints associated with a
  given value of  $\langle \sigma_y \otimes \sigma_y \rangle $,
 and the normalization of $\rho $. This variational
 problem can be cast as

\be \label{maxent1}
 \delta \Bigl[ {\rm Tr} (\rho^2) \, + \, \beta
\langle \sigma_y \otimes \sigma_y \rangle -\alpha {\rm Tr} (\rho)
\Bigr] \, = \, 0, \ee

\noindent where $\alpha $ and $\beta $ are appropriate Lagrange
multipliers. The solution of the above variational equation is
given by the density matrix

\be \label{maxent2} \rho_m \, = \, \frac{1}{2} \Bigl[ \alpha I \,
- \, \beta (\sigma_y \otimes \sigma_y ) \Bigr]. \ee

\noindent The value of the Lagrange multiplier $\alpha$ is
immediately determined by the normalization requirement,

\be \label{alph} {\rm Tr} (\rho) = 1 \,\,\, \Longrightarrow \,\,\,
\alpha = \frac{1}{2}. \ee

\noindent Consequently, those two-rebits states yielding the extremum values of
$Tr(\rho^2)$ (and also the extremum values of $R=1/Tr(\rho^2$)) compatible both
with normalization and a given value of  $\langle \sigma_y \otimes \sigma_y
\rangle $ are described by the density matrix

\be \label{maxent3}
 \rho_m \, = \, \frac{1}{4} I \,
- \, \frac{1}{2} \beta \, (\sigma_y \otimes \sigma_y ),
\ee

\noindent with the Lagrange multiplier $\beta$ lying in the
interval

\be \label{berango}
\beta \in
\left[-\frac{1}{2},\frac{1}{2}\right].
 \ee

 \noindent
 Density matrices of the form (\ref{maxent3}),
 corresponding to negative values of $\beta$,
 have been considered in \cite{CFR01}, although
 not in connection with the variational problem
 that we are discussing here.

 In terms of the parameter $\beta$, the expectation value
  $\langle \sigma_y \otimes \sigma_y \rangle $,
 the concurrence squared $C^2$, and the participation
 ratio of the statistical operator $\rho_m$ are given
 by

 \ben \label{barco2}
 \langle \sigma_y &\otimes &\sigma_y \rangle \, = \, -2\beta, \cr
 C^2 \, &=& \,   \langle \sigma_y \otimes \sigma_y \rangle^2 \, = \, 4\beta^2, \cr
  R\, &=& \, \frac{4}{1+4\beta^2},
 \een

 \noindent
 Hence, the maximum value of $R$ compatible with a given value
 of $C^2$ is given by

 \be \label{betin}
 R_m(C^2) \, = \, \frac{4}{1+C^2}.
 \ee

 \noindent
 $R_m(C^2)$ is a monotonic decreasing function of $C^2$ and adopts
 its values in the interval $2\le R\le 4$. This implies that, within
 this range of $R$-values, the maximum value of $C^2$ compatible with
 a given value of $R$ is the one obtained when solving Eq. (\ref{betin})
 for $C^2$, namely,

 \be \label{betoff}
 C^2 \, = \,  \frac{4}{R}-1.
 \ee

 \noindent
 On the other hand,
 for $1\le R \le 2$ there always exist density matrices of maximum
 entanglement (that is, with $C^2=1$). As a consequence, the maximum
 value of $C^2$ compatible with a given value of $R$
 is given by

 \be \label{cedoser}
 C^2_m \, = \, \left\{ 1 \,\,\,\,\,\,\,\,\,\,\,\,\,\, ; \,\,\, 1\le R \le 2
 \atop \frac{4}{R}-1 \,\,\,         ; \,\,\, 2\le R \le 4 \right.
 \ee

 \noindent
In Fig. 5 we plot (in the $(R,C^2)$-plane) one million numerically
generated random two-rebits states. The solid line corresponds to
the maximum concurrence squared $C^2_m$, for a given value of the
participation radio $R$, as given by Eq. (\ref{cedoser}).

\section{Conclusions}

We have explored numerically the entanglement properties of two-rebits systems.
We paid particular attention to the relationship between entanglement and
purity. We have computed numerically the mean entanglement of formation of
two-rebits systems (as determined by the CFR formula (\ref{renta1})) as a
function of the participation ratio $R$. We have also determined numerically
the probability densities $P(E)$ of finding (i) pure two-rebits states and (ii)
arbitrary two-rebits states, endowed with a given amount of entanglement $E$.
Furthermore, we surveyed the distribution of general two-rebits states in the
$(R,C^2)$-plane. In particular, we determined analytically the maximum possible
value of the concurrence squared $C^2$ of two-rebits states compatible with a
given value of the participation ratio. An interesting feature that deserves
special mention is that, with regards to the probability of finding states with
a given amount of entanglement, the difference between mixed states and pure
states is much larger for qubits than for rebits.

 It would be interesting to perform, for quaternionic quantum mechanics
\cite{A95}, an analysis similar as the one done here.

\acknowledgments This work was partially supported by the  DGES grants
PB98-0124  (Spain), and by CONICET (Argentine Agency).

\newpage

\noindent {\bf FIGURE CAPTIONS}

\vskip 0.5cm

\noindent Fig. 1- Mean entanglement of formation $\langle E
\rangle $ of real density matrices as a function of the
participation ratio $R$. The continuous line corresponds to the
CFR formula. The dashed line corresponds to the mean entanglement
of the same matrices, as given by Wootters' formula.

\vskip 0.5cm

\noindent  Fig. 2- The continuous line shows the behavior of the mean
entanglement of formation $\langle E \rangle $  of real density matrices (given
by the CFR expression) as a function of the participation ratio $R$. The dashed
line shows the behavior of the mean entanglement of formation $\langle E
\rangle $ of complex density matrices (given by Wootters formula) as a function
of the participation ratio $R$.

\vskip 0.5cm

\noindent Fig. 3- Mean entanglement $\langle E\rangle$ of all the two-rebits
states with a given value of their maximum eigenvalue $\lambda_m$, as a
function of this last quantity. The upper line corresponds to the CFR
expression and the lower line to Wootters formula. Notice that, in the case of
Wootters formula, the mean entanglement vanishes for $\lambda_m \le 1/3 $.

\vskip 0.5cm

\noindent  Fig. 4a- Plot of the probability $P(E)$ of finding
two-qubits states endowed with a given entanglement $E$. The solid
line correspond to arbitrary states and the dashed line to pure
states.

\vskip 0.5cm

\noindent  Fig. 4b- Plot of the probability $P(E)$ of finding two-rebits states
endowed with a given entanglement $E$. The solid line correspond to arbitrary
states and the dashed line to pure states. The horizontal line corresponds to
the limit value $P(E=1)=\ln 2$ of the probability density  associated with pure
two-rebits states.

\vskip 0.5cm

\noindent  Fig. 5- Plot of in the $(R,E)$-plane of one million
random numerically generated two-rebits states. The solid line
corresponds to the maximum entanglement $E_m$, for a given value
of the participation radio $R$.

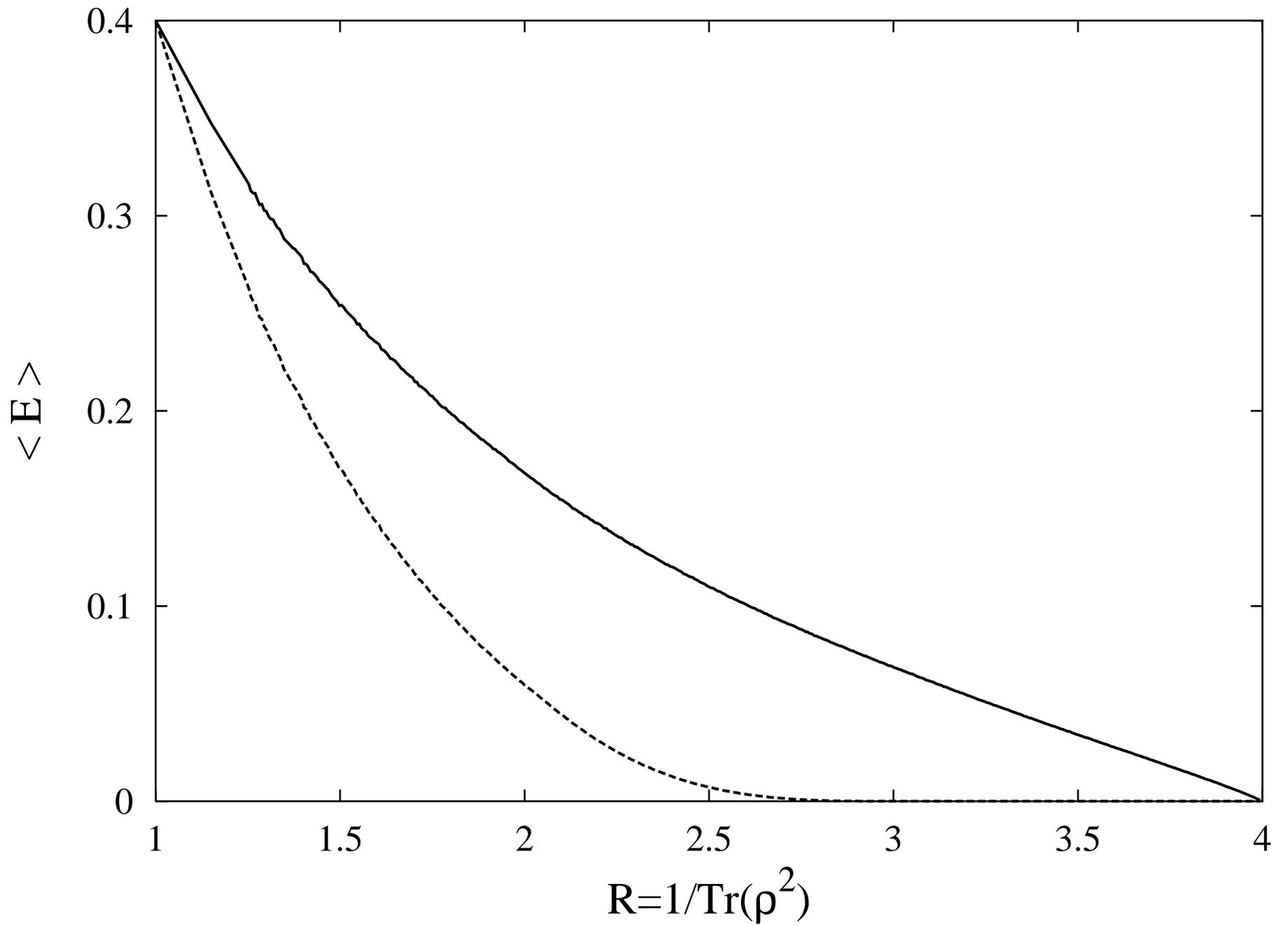
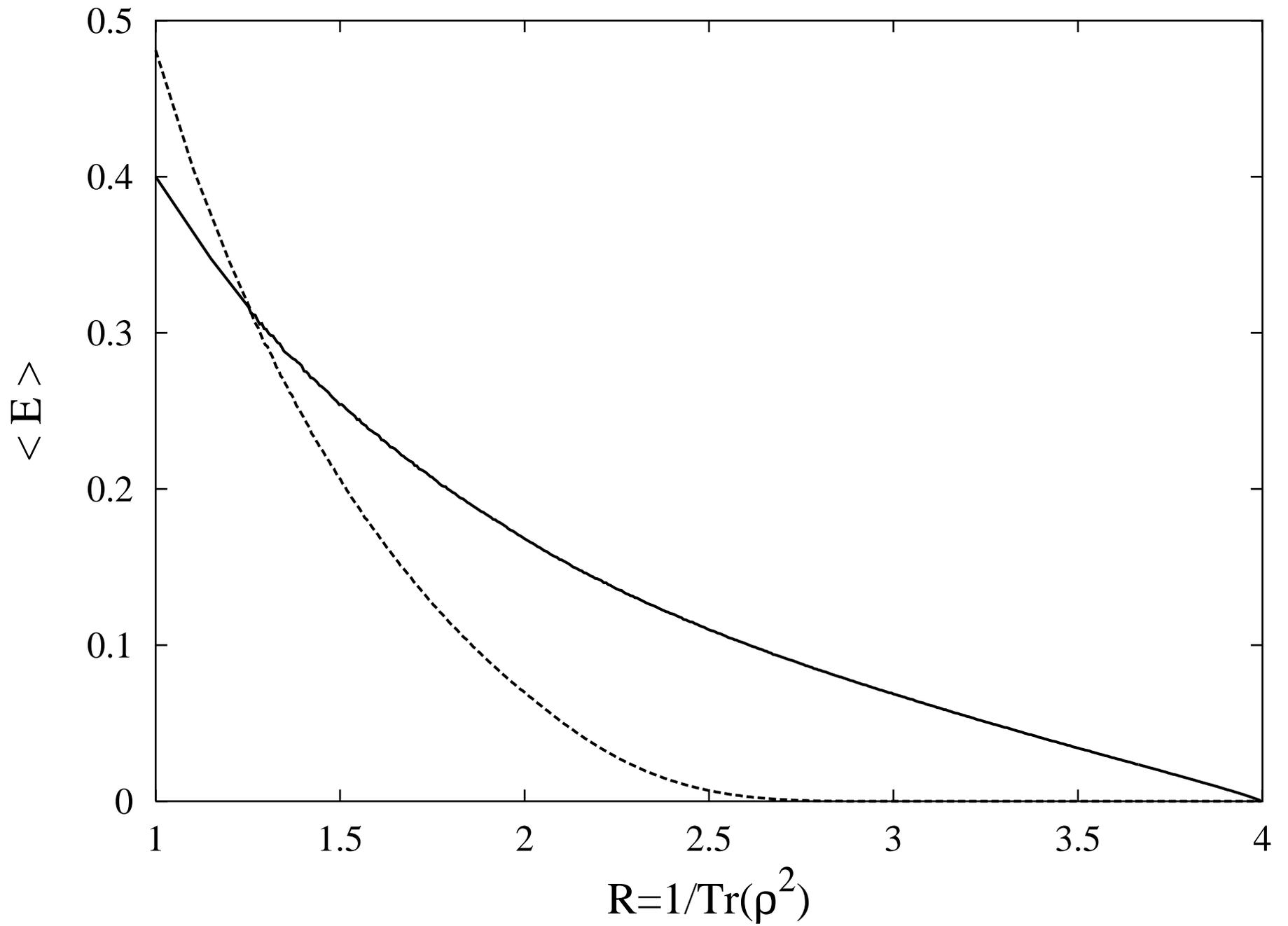
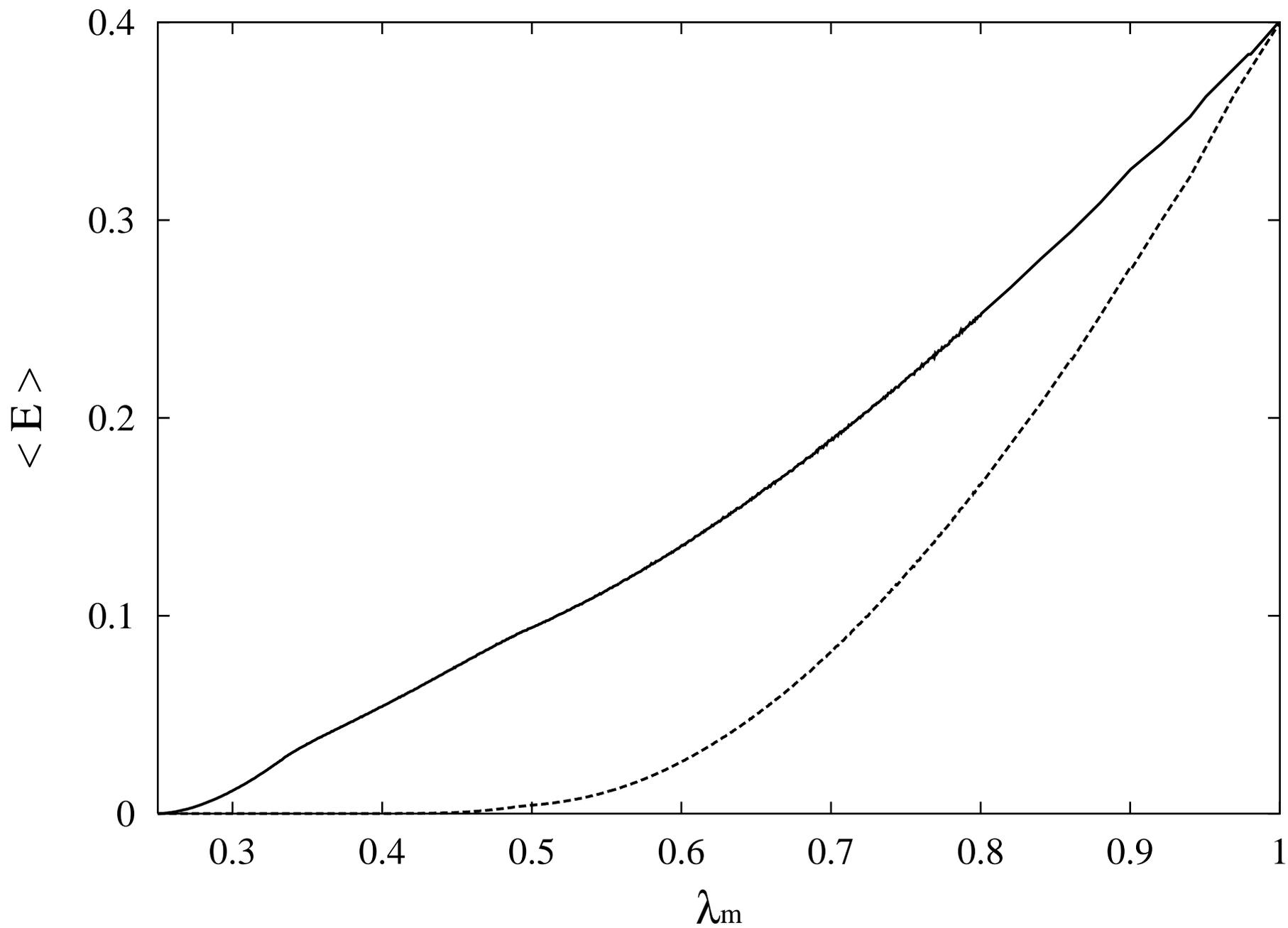
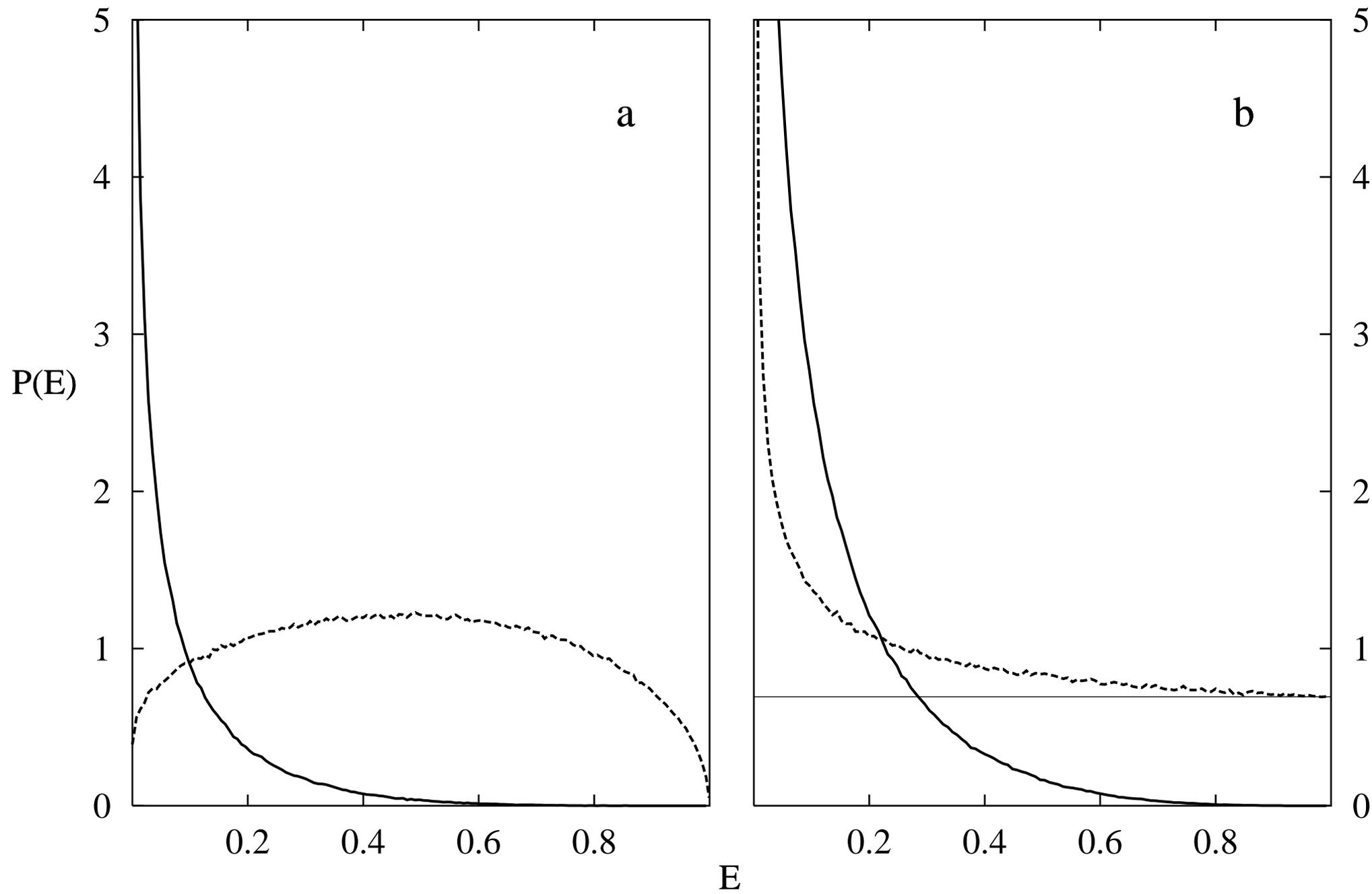
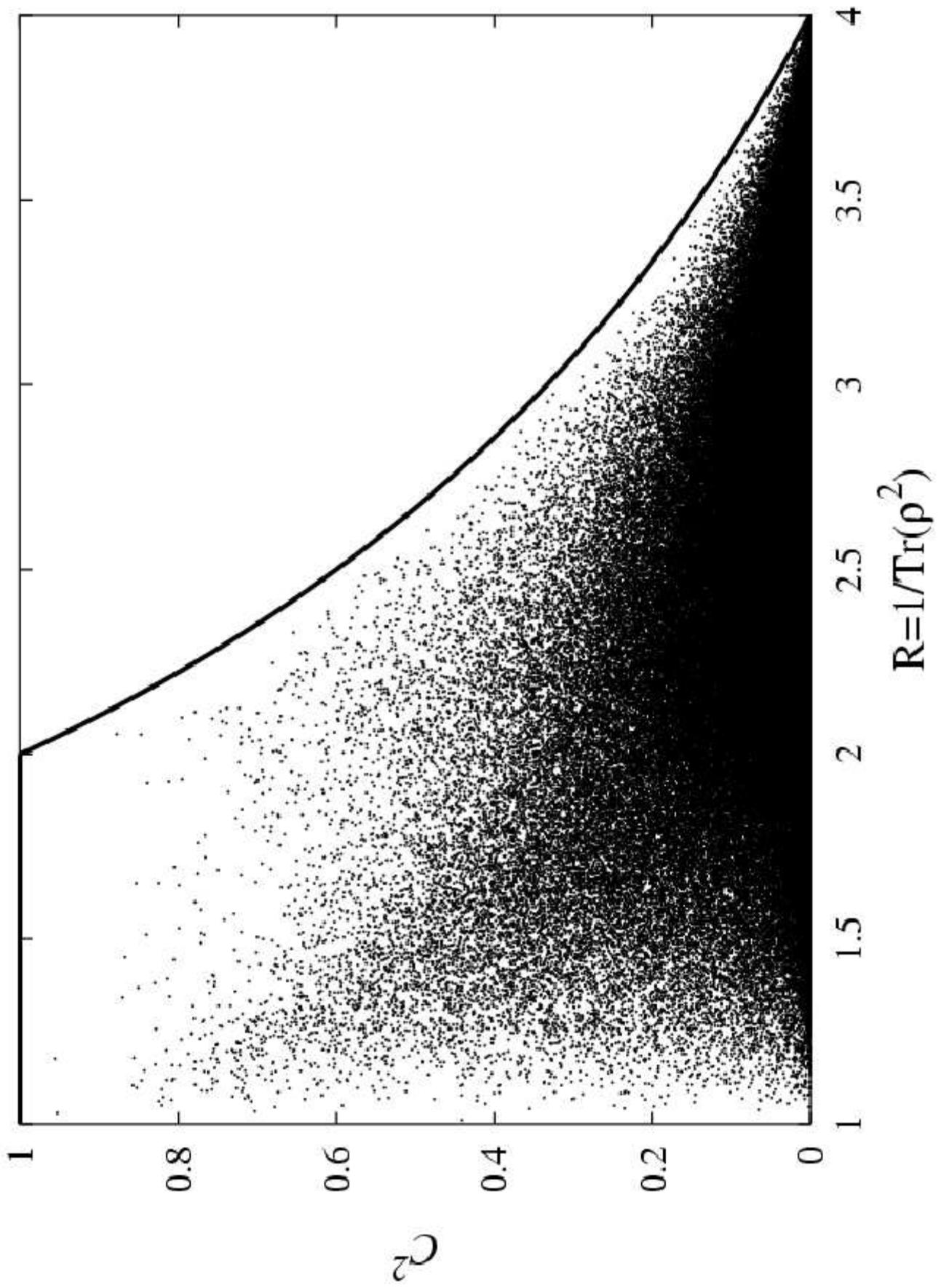
\end{document}